\documentclass{emulateapj}

\usepackage{amsmath}

\def\H2{{{\rm H}_2}}

\shorttitle{On the time variability of the star formation efficiency}
\shortauthors{Feldmann \& Gnedin}

\begin{document}


\title{On the time variability of the star formation efficiency}


\author{R. Feldmann\altaffilmark{1,2} and N. Y. Gnedin\altaffilmark{1, 2, 3}}


\altaffiltext{1}{Particle Astrophysics Center, 
Fermi National Accelerator Laboratory, Batavia, IL 60510, USA; feldmann@fnal.gov}
\altaffiltext{2}{Kavli Institute for Cosmological Physics and Enrico
  Fermi Institute, The University of Chicago, Chicago, IL 60637 USA} 
\altaffiltext{3}{Department of Astronomy \& Astrophysics, The
  University of Chicago, Chicago, IL 60637 USA} 


\begin{abstract}
A star formation efficiency per free fall time that evolves over the 
life time of giant molecular clouds (GMCs) may have important implications for models of supersonic turbulence in molecular clouds or for the relation
between star formation rate and $\H2$ surface density.
We discuss observational data that could be interpreted as evidence of such a time variability. In particular, we investigate a recent claim based on
measurements of $\H2$ and stellar masses in individual GMCs. We show that this claim depends crucially on the assumption that $\H2$ masses do
not evolve over the life times of GMCs. We exemplify our findings with a simple toy model that  uses a constant star formation efficiency and, yet,
is able to explain the observational data. 
\end{abstract}





\section{Introduction}
\label{sect:intro}

The life times of giant molecular clouds (GMCs) have been at the center of a major debate for at least the last 40 years \citep{1974ApJ...189..441G, 1974ApJ...192L.149Z, 1979IAUS...84...35S, 2000ApJ...530..277E,2007ApJ...668.1064E}. GMCs that live for many free-fall times need a mechanism that prevents them from gravitational collapse. Over the last couple of years the consensus is growing that the life times of GMCs are likely a few free-fall times, or even less \citep{2000ApJ...530..277E, 2007RMxAA..43..123B, 2010arXiv1007.3270M} and the focus has shifted towards the challenge of explaining the low star formation efficiencies in GMCs. 
The star formation efficiency per free fall time $\epsilon_{\rm ff}$ is defined as the ratio of free fall time $t_{\rm ff}$ to gas depletion time $M_{\H2}/\dot{M}_*$. In other words:
\begin{equation}\label{eq:dotMstar}
\dot{M}_*=\frac{\epsilon_{\rm ff}}{t_{\rm ff}}M_{\H2},
\end{equation}
i.e. the instantaneous star formation rate (SFR) is proportional to the available amount of molecular hydrogen ($\H2$) via the proportionality factor $\epsilon_{\rm ff}/t_{\rm ff}$.
The observed value $\epsilon_{\rm ff}\sim{}0.01-0.02$ (e.g.  \citealt{2007ApJ...654..304K}) means that only 1-2\% of the mass of a GMC is converted into stars
over a free-fall time. If star formation is supported by supersonic turbulence \citep{2005ApJ...630..250K} $\epsilon_{\rm ff}$ is expected to be only very weakly dependent on the Mach number of the turbulent flow and thus approximatively constant, but this may be an oversimplification, see e.g. \citep{2005ApJ...630L..49V, 2006ApJ...640L.187L}. On the other hand, if GMCs have life times of the order of a free fall time they do not need be supported by turbulence. The star formation efficiency in such clouds may increase as the Mach number in the flow decreases and the cloud collapses, see e.g. \citealt{2010arXiv1009.1152B}. We note that a time varying $\epsilon_{\rm ff}$ should introduce additional scatter in the relation between star formation rate and $\H2$ surface density on small ($\lesssim{}100$ pc) scales. This scatter should propagate up to $\sim{}$kpc scales, see e.g. \citep{Feldmann0}, and hence would contribute to the scatter in the Kennicutt-Schmidt relation. This, at least in principle, could be used to test observationally the time dependence of $\epsilon_{\rm ff}$.

In section \ref{sect:pitfalls} we will discuss two common misconceptions that could give rise to the impression that $\epsilon_{\rm ff}$ varies over the life times of GMCs even if it is a constant. In section \ref{sect:model} we present and analyze a toy model in order to exemplify and quantify our statements.

\section{Do observations confirm a time-varying star formation efficiency?}
\label{sect:pitfalls}

In equation (\ref{eq:dotMstar}) we define the \emph{star formation efficiency per free fall time} $\epsilon_{\rm ff}$.
Another commonly used efficiency is the \emph{star formation efficiency of the GMC} $\epsilon_{\rm GMC}$, i.e. the fraction of $\H2$ mass of the cloud that is converted into stars over the life time of the cloud. In a picture where GMCs start with an initial reservoir of $\H2$, which is used in the subsequent star formation process, the final stellar mass $M_*({\rm final})$ is divided by the \emph{initial} $\H2$ mass of the cloud. If the cloud accretes a substantial amount of $\H2$ over its life time, the definition has to be generalized. We will use: 
\begin{equation}\label{eq:GMCeps}
\epsilon_{\rm GMC} = \frac{M_*({\rm final})}{\max(M_\H2)},
\end{equation}
where $\max(M_\H2)$ is the \emph{maximal} $\H2$ mass of the GMC. By definition $\epsilon_{\rm GMC}$ is a non evolving quantity and it can be estimated, e.g., by comparing luminosity distribution of OB associations in the Milky Way with the mass spectrum of molecular clouds \citep{1997ApJ...476..166W}. It \emph{cannot} be directly measured on a cloud-to-cloud basis, because $M_*$ and $M_\H2$ must be known at two different times. Instead such observations, see e.g. \citep{1986ApJ...301..398M}, estimate the following quantity
\begin{equation}\label{eq:GMCeta}
\eta_{\rm GMC}(t) =  \frac{M_*(t)}{M_\H2(t)+M_*(t)}\approx{}\frac{M_*(t)}{M_\H2(t)}.
\end{equation}
The latter, approximate equality is due to the fact that for most observed GMCs $M_*$ is smaller than $M_\H2$. Obviously, $\eta_{\rm GMC}(t)$ increases over the life time of a cloud and should not be confused with either $\epsilon_{\rm GMC}$ or $\epsilon_{\rm ff}$.
From (\ref{eq:dotMstar}) we can estimate $M_*({\rm final})=\xi\epsilon_{\rm ff}/t_{\rm ff}\max(M_\H2)t_{\rm final}$, hence $\epsilon_{\rm GMC}=\xi\epsilon_{\rm ff}t_{\rm final}/t_{\rm ff}$, 
where $\xi$ is a constant fudge factor of order unity that depends on the actual time evolution of the SFRs and $M_\H2$ (and $\epsilon_{\rm ff}$ if it is time dependent). We will estimate $\xi$ for a simple toy model in section \ref{sect:model}.
Combining this result with equation (\ref{eq:GMCeps}) and (\ref{eq:GMCeta}) we obtain:
\begin{equation}\label{eq:GMCeps2}
\eta_{\rm GMC}(t) \approx{} \xi\epsilon_{\rm ff} \left[\frac{t_{\rm final}}{t_{\rm ff}}\right] \left[\frac{M_*(t)}{M_*({\rm final})}\right]\left[\frac{\max(M_\H2)}{M_\H2(t)}\right].
\end{equation}
There are several ways of creating large values of $\eta_{\rm GMC}$ and they correspond to the various terms in this equation. 
First, $\epsilon_{\rm ff}$ could be time dependent. For instance, it could smoothly increase  
as the cloud collapse advances or, alternatively, vary stochastically about some average value. 
A second possibility is that some clouds may live for many free fall times, i.e. $t_{\rm final}/t_{\rm ff}$ is large in a subset of GMCs. The third factor in the third bracket in equation (\ref{eq:GMCeps2}) explain why $\eta_{\rm GMC}$ can also be \emph{smaller} than $\epsilon_{\rm GMC}$. Finally, $\eta_{\rm GMC}$ can be boosted if the observed $\H2$ mass is significantly less than $\max(M_\H2)$, i.e. if GMCs lose (in one way or another) a large fraction of their molecular hydrogen over their life time.
The latter scenario predicts that $\eta_{\rm GMC}(t)$ should roughly scale $\propto{}M_\H2^{-1}$ over the life time of \emph{individual} GMCs. An observational sample of an \emph{ensemble} of GMCs shows this trend \citep{2010arXiv1007.3270M}. However, this trend can also be produced by a selection effect based on stellar mass, e.g. selecting GMCs with $M_* > M_{*,{\rm limit}}$ excludes values of $\eta_{\rm GMC}$ that are smaller than $M_{*,{\rm limit}}/M_\H2$, see equation (\ref{eq:GMCeta}). 
In fact, \cite{2010arXiv1007.3270M} is selecting clouds based on ionizing luminosities, which roughly corresponds to selecting clouds based the stellar mass formed within the last 4 Myr. Such a selection effect explains why a different study of $\sim{}10^5 M_\odot$ GMCs find much lower efficiencies \cite{2010arXiv1009.2985L}. The existence of the selection effect is \emph{not} an argument against or in favor of an evolving $\epsilon_{\rm ff}$, rather it shows that the GMCs with large values of $\eta_{\rm GMC}$ in the sample of \cite{2010arXiv1007.3270M} are likely a heavily biased subset. 
In the case that $\epsilon_{\rm ff}$ is, in fact, a non evolving quantity and the measured large values of  $\eta_{\rm GMC}$ are driven by changing molecular gas masses, we can make a rather generic prediction. The similarity of the scaling with GMC mass\footnote{A linear regression of $\eta_{\rm GMC}$ vs. $M_\H2$ for the data presented in \cite{2010arXiv1007.3270M} gives a slope of $-0.59\pm{0.19}$. This is consistent with the prediction of our toy model (slope $\sim{}-0.75$, see section \ref{sect:model}) that takes into account that, in fact, not the total stellar mass has been measured, but only the stellar mass formed within the last $\sim{}$ 4 Myr.} ($\propto{}M_\H2^{-1}$) of $\eta_{\rm GMC}$, on the one hand, and the lower boundary of the region excluded by the discussed selection effect, on the other hand, implies that the observed GMCs with large values of $\eta_{\rm GMC}$ should have rather similar \emph{maximal} $\H2$ masses $\max(M_\H2)$. The toy model that we discuss in section \ref{sect:model} predicts $\max(M_\H2)\sim{}10^6-10^7 M_\odot$. We note that this scenario explains rather naturally the absence of massive ($\gtrsim{}10^6$ $M_\odot$) GMCs with high values of $\eta_{\rm GMC}$.

A different issue can arise if one compares star formation rates and $\H2$ masses in order to estimate $\epsilon_{\rm ff}/{t_{\rm ff}}$ via equation (\ref{eq:dotMstar}).
For example, let us assume that we measure SFRs and $\H2$ masses within small ($\lesssim{}100$ pc) apertures around peaks of CO emission (tracing the $\H2$ mass) and peaks of $H\alpha$ emission (tracing star formation rates), see e.g. \cite{2010arXiv1009.1651S}. If we observe that CO peaks have lower SFRs at given $\H2$ mass compared with peaks of $H\alpha$ emission, does this imply a time-varying $\epsilon_{\rm ff}/{t_{\rm ff}}$? The answer to that question depends on the way the SFRs are measured. SFRs that are derived from $H\alpha$ emission are effectively averaged over the past 5-10 Myr, which might well be a significant fraction of the life time of the molecular cloud. For SFRs that are based on $H\alpha$+$24\mu{}m$ emission this averaging time span would be even longer. The star formation efficiencies per free-fall time that are estimated from such a time averaged SFR will be small initially (no stars have been formed over most of the time averaging interval simply because the GMC has only formed recently). The measured SFRs will increase until the age of the GMC is similar to the averaging time span. In addition, the $\H2$ mass of the cloud might evolve (possibly decrease) leading to an additional increase in the apparent value of $\epsilon_{\rm ff}/{t_{\rm ff}}$ with time. If the following three conditions are satisfied, a difference in the measured SFR per measured $\H2$ mass can provide strong evidence for a time-varying star formation efficiency per free fall time.
First, the averaging times of the SFRs need to be small compared to ages of the observed clouds. Second, the observable $\H2$ reservoirs need to be close to $\max(M_\H2)$, and, finally, the free fall times of the clouds need to be known. A recent study that measures SFRs with reasonably short averaging times (2 Myr, \citealt{2010arXiv1009.2985L})
estimates star formation efficiencies per free fall time of the order of $2\%$ for most clouds in the sample, with the scatter mostly driven by the mass of molecular gas of relatively low density ($n<10^4$ cm$^{-3}$) that does not participate in the star formation.

\section{Toy Model}
\label{sect:model}

We will now discuss a toy model in order to both exemplify the points made in section \ref{sect:pitfalls}, but also to provide a framework in which we can make some quantitative predictions. We should stress that the statements made in the previous section are completely generic and do not depend on the specific assumption that go into the model that we are going to present. Our model is almost insultingly simple, and, given that, our aim is not to reproduce the full complexity in the evolution of GMCs or even, to be consistent with any available observation. On the other hand the model offers a pragmatic approach to the mass evolution of GMCs and may be easily generalized to facilitate more complex scenarios.

The ansatz of the model is to supplement equation (\ref{eq:dotMstar}) with an equivalent equation that describes the evolution of 
the $\H2$ mass:
\begin{equation}\label{eq:dotMgas}
\dot{M}_\H2 = -\frac{\epsilon_{\rm ff}}{t_{\rm ff}}M_{\H2} - \alpha{}M_* + \gamma
\end{equation}
The extra term $\alpha{}M_*$ is motivated by assuming that stellar feedback is limiting the life time of molecular clouds, e.g. via photo-ionization, thermal pressure or radiation pressure \citep{1997ApJ...476..166W, 2010ApJ...709..191M, 2010arXiv1008.2383L}. This feedback should therefore couple to the formed stellar mass via some efficiency factor $\alpha$ that sets the time scale for the destruction/removal of $\H2$ from the cloud\footnote{Depending on the type of feedback $M_*$ should refer to the total stellar mass times a weight parameter that takes into account that feedback is provided by stars which have a limited life time. For simplicity we will assume that $M_*$ is the total amount of stellar mass formed within the cloud.}.
The term $\gamma$ is the net ``accretion'' rate of $\H2$, which includes all processes that create and destroy $\H2$ and are not directly coupled to either $M_*$ or $M_\H2$. Both $\alpha$ and $\gamma$ could in principle be time dependent. For simplicity we assume that they are constant. Our model is minimalistic (compared with, e.g., \citealt{2002ApJ...566..302M, 2006ApJ...641L.121T, 2006ApJ...644..355H, 2006ApJ...653..361K}), but it has the advantage that we can parametrize our ignorance of the relevant physical processes that destroy and disperse the cloud into the parameters $\alpha$ and $\gamma$. Together with appropriate initial conditions equations (\ref{eq:dotMstar}) and (\ref{eq:dotMgas}) fully determine the evolution of the masses of molecular hydrogen and the stellar component in a GMC.

We will also make the simplifying assumption that the free-fall time does not evolve strongly over the history
of the GMC, i.e. both the star formation efficiency per free fall time and the star formation time scale are now fixed.
This assumption is not crucial for the model, but we will use it for the following reasons. First, there is no clear systematic trend of free fall time with mass over the range of GMCs that we are comparing to, see e.g. Table 2 of \cite{2010arXiv1007.3270M}. Second, assuming a non-evolving free-fall time allows for a convenient analytical solution of the problem. Third, we find that even with this assumption our model describes the observed data reasonably well. We stress that our main aim 
is to show that a simple model can produce an observational signal that could be misinterpreted as evidence for evolution of the star formation efficiencies. We do not try to model the precise properties of the ensemble of GMCs in the Galaxy.

With $t_{\rm ff}$ fixed (and, of course, we assume that the  star formation efficiency per free fall time is a constant, too) we can insert  (\ref{eq:dotMstar}) into (\ref{eq:dotMgas}) and obtain a linear 2nd order differential equation for $M_\H2$, i.e. the equation of a damped harmonic oscillator.

Solving the differential equation we obtain
\begin{eqnarray}
M_\H2(t) &=& Ae^{-tb/2}\cos(\omega{}t+\phi), \label{eq:MH2} \\
M_*(t) & = &\frac{M_\H2}{\alpha} ( \omega \tan(\omega{}t+\phi) - b/2)+\frac{\gamma}{\alpha}, \label{eq:Mstar}
\end{eqnarray}
where  $b=\epsilon_{\rm ff}/t_{\rm ff}$ is the inverse of the star formation timescale, and
$\omega = \sqrt{\alpha{}b- b^2/4}$ is the ``oscillation'' period.

Phase $\phi$ and amplitude $A$ depend on the initial conditions. In the following we restrict ourselves to two special cases of the general model (\ref{eq:MH2}), (\ref{eq:Mstar}).
\begin{itemize}
\item \emph{No accretion scenario:}  Assumes $\gamma=0$, $M_\H2(t=0)=M_0>0$, and $M_*(t=0)=0$. It follows $\phi={\rm atan}(b/(2\omega))$, and $A  =  M_0/\cos(\phi)$.
\item \emph{Pure accretion scenario:} Assumes that all $\H2$ is ``accreted'', i.e. $M_\H2(t=0)=0$, $M_*(t=0)=0$ and $\gamma>0$. In this case phase and amplitude are given by $\phi=-\pi/2$, $A=\gamma/\omega$.
\end{itemize}

We adopt the parameters $\epsilon_{\rm ff}=0.02$ and $t_{\rm ff}=6$ Myr, which are consistent with observations of $\epsilon_{\rm ff}$ over a range of density scales \citep{2007ApJ...654..304K}, and with the free fall times $6.1^{+6.8}_{-4.0}$ Myr measured in the sample of \cite{2010arXiv1007.3270M}, respectively. We note that only the ratio $\epsilon_{\rm ff}/t_{\rm ff}=0.0033$ Myr$^{-1}$ enters our model.  The $\alpha$ parameter is chosen such that the life time of the cloud, i.e. the time $t_{\rm final}$ at which $M_\H2(t_{\rm final})=0$, is $\sim{}20$ Myr \citep{1997ApJ...476..166W}. Hence, we use $\alpha = 2 {\rm\, Myr}^{-1}$ in the no accretion scenario and $\alpha = 8 {\rm\, Myr}^{-1}$ in the pure accretion scenario, respectively.
 
\begin{figure*}
\epsscale{.48}
\begin{tabular}{cc}
\includegraphics[width=85mm]{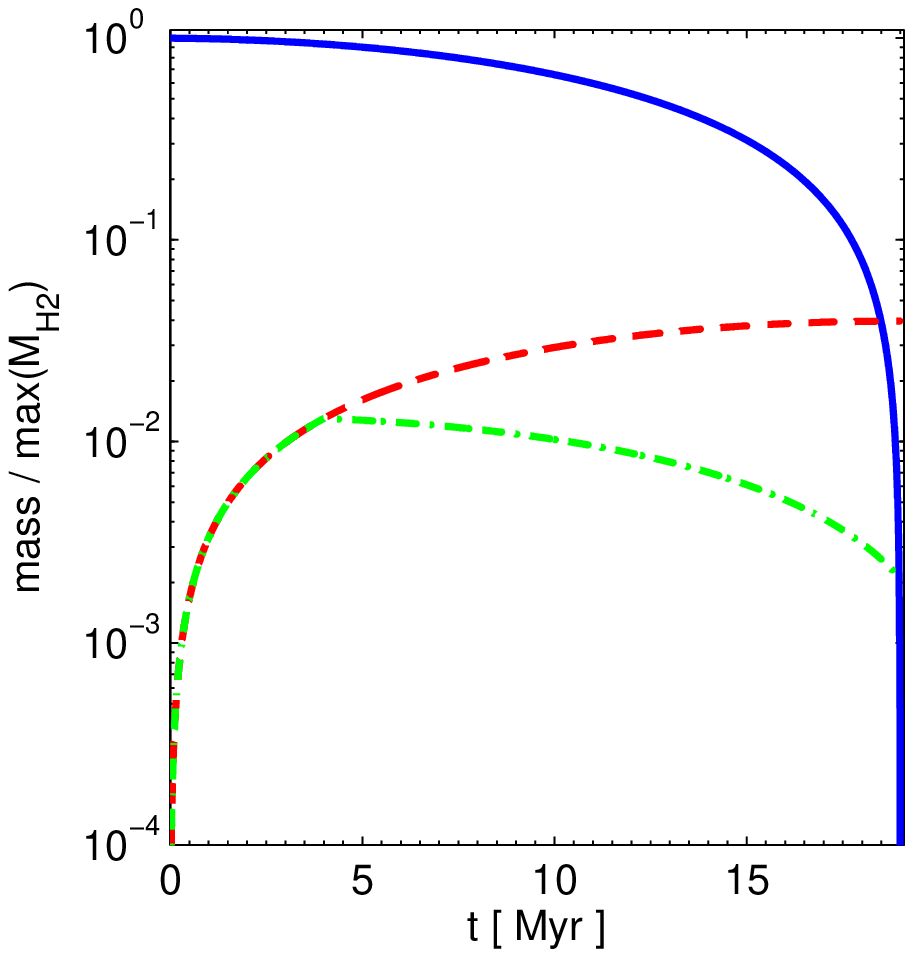} &  \includegraphics[width=85mm]{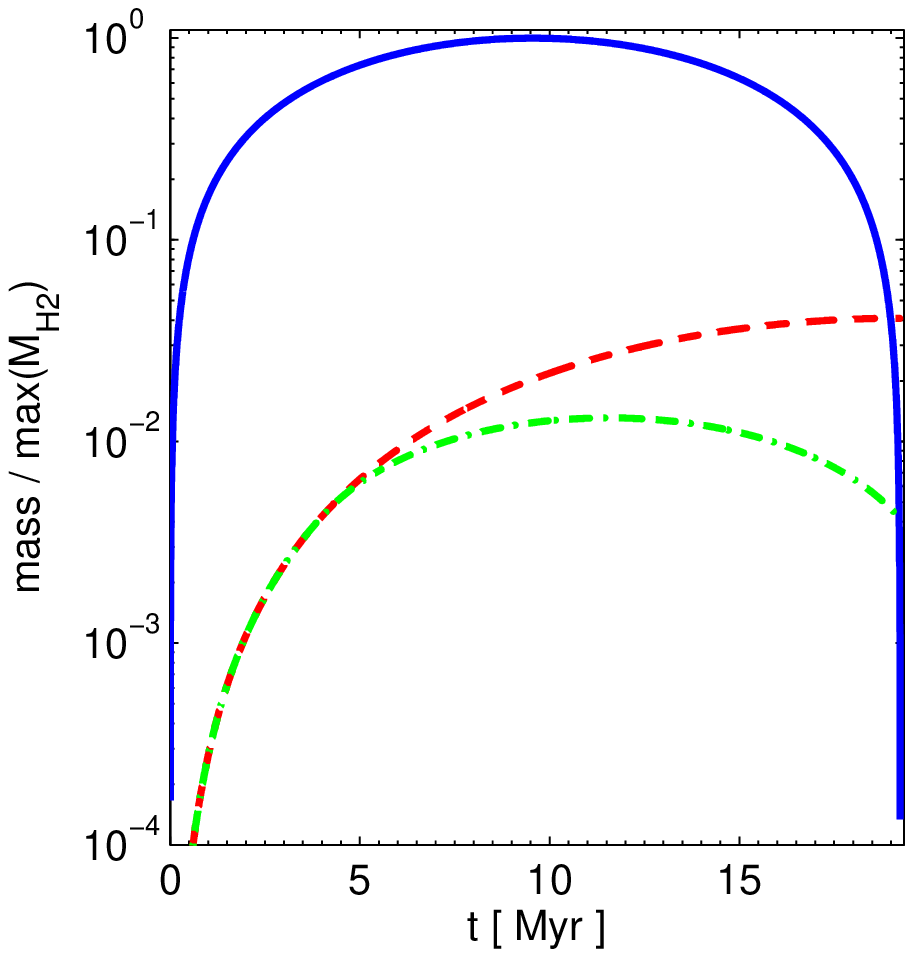}
\end{tabular}
\caption{The evolution of the masses of the GMC components (normalized to the maximum $\H2$ mass of the GMC) according to the two scenarios: \emph{no accretion} (left) and \emph{pure accretion} (right). We assume $\epsilon_{\rm ff}/t_{\rm ff}=0.0033 {\rm\, Myr}^{-1}$ and $\alpha= 2{\rm\, Myr}^{-1}$ ($\alpha= 8{\rm\, Myr}^{-1}$) in the no accretion (pure accretion) scenario. The different lines correspond to the $\H2$ mass (solid blue line), total stellar mass (dashed red line), stellar mass formed within 4 Myr (dot-dashed green line).\label{fig:tM}}
\end{figure*}

Assuming $\epsilon_{\rm ff}/t_{\rm ff}\ll{}\alpha$ the life time of a GMC is given by
\[
t_{\rm final}\approx{}\frac{\pi}{2\sqrt{\alpha \epsilon_{\rm ff}/t_{\rm ff}}},\textrm{ and }t_{\rm final}\approx\frac{\pi}{\sqrt{\alpha \epsilon_{\rm ff}/t_{\rm ff}}}.
\]
The left (right) expression refers to the no accretion (pure accretion) scenario. We note that in both considered scenarios the life time does not depend on the initial cloud mass or the accretion rate, respectively.  The evolution of $M_\H2$, $M_*$ and $M_{*, <4 {\rm Myr}}$ normalized to $\max(M_\H2)$ is shown in Fig.~\ref{fig:tM}.

Assuming $\epsilon_{\rm ff}/t_{\rm ff}\ll{}\alpha$ we can easily estimate the total stellar mass that is formed during the life time of the cloud from equation (\ref{eq:Mstar}).
In the no accretion scenario we obtain
\[
M_*(t_{\rm final}) \approx{}  M_0 \sqrt{\frac{\epsilon_{\rm ff}/t_{\rm ff}}{\alpha}}\left[ 1 -  t_{\rm final} \frac{  \epsilon_{\rm ff}  }{ 2t_{\rm ff } } \right] \approx{} M_0 \sqrt{\frac{\epsilon_{\rm ff}/t_{\rm ff}}{\alpha}},
\]
while the pure accretion scenario predicts
\[
M_*(t_{\rm final}) \approx{} \frac{2\gamma}{\alpha}\left[ 1 -  t_{\rm final} \frac{  \epsilon_{\rm ff}  }{ 2t_{\rm ff } } \right] \approx{} \frac{2\gamma}{\alpha}.
\]
In the pure accretion scenario a GMC attains its maximum mass at $t\approx{}t_{\rm final}/2$. The $\H2$ mass is then approximatively $\gamma/\sqrt{\alpha{}\epsilon_{\rm ff}/t_{\rm ff}}$. Combining these results we see that the star formation efficiency of a GMC is
\[
\epsilon_{\rm GMC}\approx{} \sqrt{\frac{\epsilon_{\rm ff}/t_{\rm ff}}{\alpha}},\,\textrm{and } \epsilon_{\rm GMC} \approx{} 2\sqrt{\frac{\epsilon_{\rm ff}/t_{\rm ff}}{\alpha}}.
\]
Again, the left (right) expression refers to the no accretion (pure accretion) scenario.
Written in terms of the life time of the GMC both expression are identical, namely $\epsilon_{\rm GMC}/t_{\rm final} \approx{} (2/\pi) \, \epsilon_{\rm ff}/t_{\rm ff}$, i.e. $\xi=2/\pi$ (section \ref{sect:pitfalls}).

\begin{figure*}
\epsscale{.48}
\begin{tabular}{cc}
\includegraphics[width=85mm]{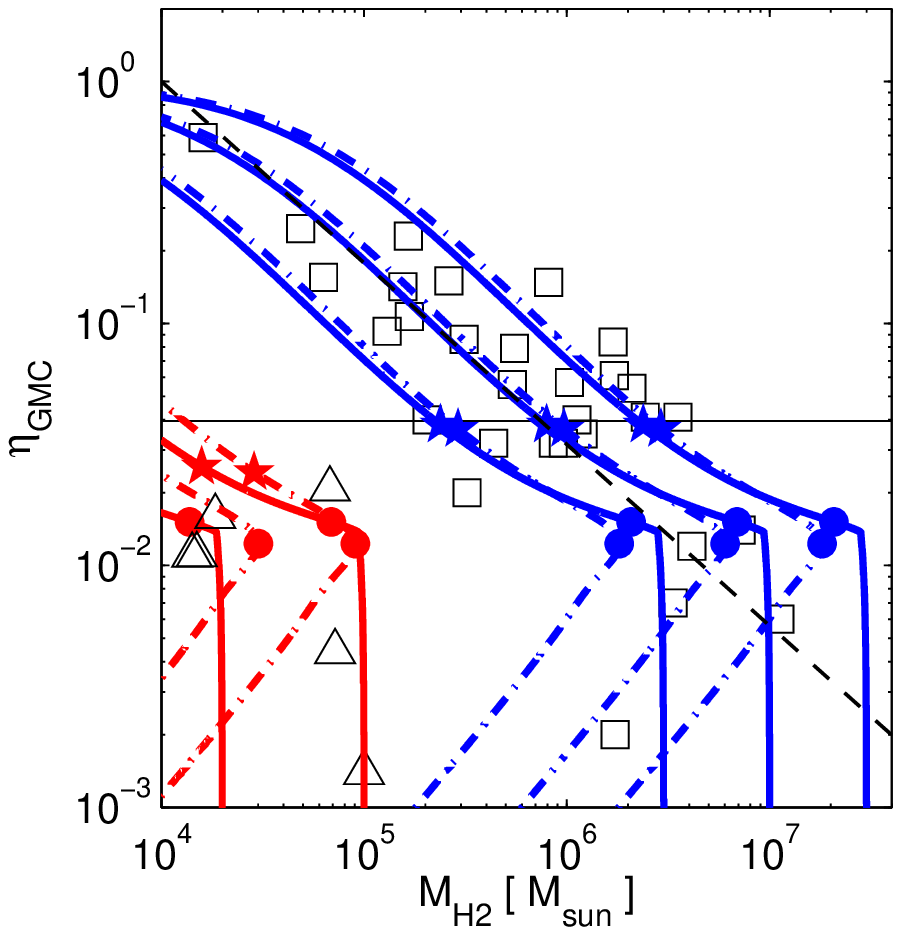} &  \includegraphics[width=85mm]{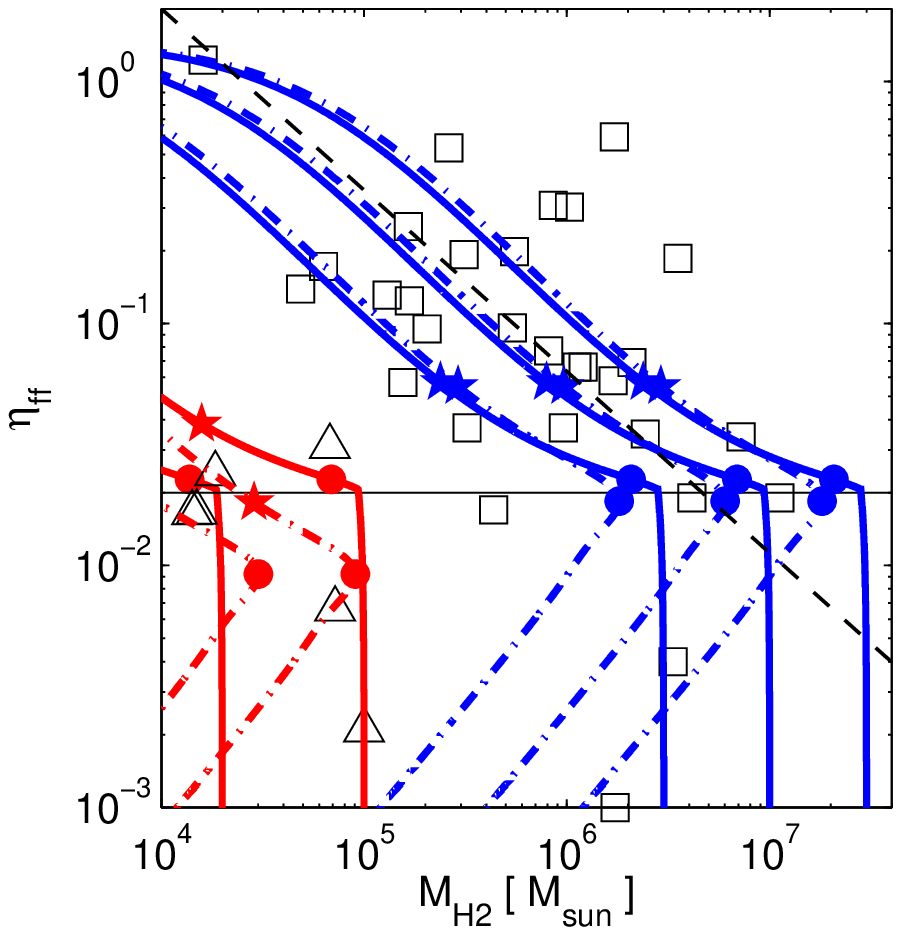}
\end{tabular}
\caption{Estimators of the star formation efficiencies as function of molecular mass of the GMC. The quantity $\eta_{\rm GMC}$ ($\eta_{\rm ff}$) is shown in the left (right) panel. The observational data presented by \cite{2010arXiv1007.3270M} and \cite{2010arXiv1009.2985L} are indicated with empty squares and triangles, respectively. 
Overplotted are solid (dot-dashed) lines that refer to the predictions of the toy model in the no accretion (pure accretion) scenario. The 6 solid and dot-dashed blue lines at the top
use $\epsilon_{\rm ff}/t_{\rm ff}=0.0033 {\rm\, Myr}^{-1}$, $\alpha= 2{\rm\, Myr}^{-1}$ ($\alpha= 8{\rm\, Myr}^{-1}$) in the no accretion (pure accretion) scenario and should be compared to the squares. Note that for consistency only the stellar mass formed within the last 4 Myr is considered in the computation of $\eta_{\rm GMC}$ and $\eta_{\rm ff}$. Red lines (near the bottom left of each panel) use free fall times that are a factor 2 smaller (since these are intrinsically smaller clouds), $\alpha$ that are a factor 2 larger (to keep the same $\epsilon_{\rm GMC}$) and the stellar mass that formed within the past 2 Myr (the masses are derived from counting young stellar objects). These lines should be compared to the triangles. Filled circles and filled stars indicate when the age of the modeled GMC is half its total life time or when the cloud is 1 Myr away from the end of its life, respectively. The diagonal dashed line indicates a slope of -0.75, which is approximatively the slope predicted by our toy model. The observed data is consistent with this slope. A linear regression of $\eta_{\rm GMC}$ and $\eta_{\rm ff}$ as function of GMC mass using all clouds with masses $>10^4 M_\odot$ returns slopes of $-0.59\pm{0.19}$ and $-0.49\pm{0.32}$, respectively, at the 95\% confidence limit. Each panel also contains a horizontal line that denotes the value of the star formation efficiencies $\epsilon_{\rm ff}=0.02$ and $\epsilon_{\rm GMC}=0.04$, respectively. \label{fig:eps}}
\end{figure*}

In Fig. \ref{fig:eps} we show the predictions for $\eta_{\rm GMC}$ and $\eta_{\rm ff}$ of the two scenarios of our model, together with  $\epsilon_{\rm ff}$ and $\epsilon_{\rm GMC}$, and the observational data from \cite{2010arXiv1007.3270M}. To be consistent with observations only the stellar mass that formed within the last 4 Myr is included in the definition $\eta_{\rm GMC}$, see (\ref{eq:GMCeta}), and $\eta_{\rm ff}$ is estimated as $\eta_{\rm ff} = \eta_{\rm GMC}\, t_{\rm ff}/4\,{\rm Myr}$.
Our model reproduces the trends of $\eta_{\rm GMC}$ and $\eta_{\rm ff}$ with GMC mass, suggesting that these are maybe not solely due to selection effects.
In both scenarios  $\eta_{\rm GMC}$ and $\eta_{\rm ff}$ roughly scale as $M_\H2^{-0.75}$ over the mass range $10^4 M_\odot$ - $10^6 M_\odot$. 

With the chosen parameters our model predicts that $\eta_{\rm GMC}$ is only  
significantly larger than $\epsilon_{\rm GMC}$ for the last $\sim{}1$ Myr in the life of a GMC, this includes most of the
GMCs with masses less than $\sim{}10^6 M_\odot$ in the sample of \cite{2010arXiv1007.3270M}. We note that the precise
time does depend on the assumed life time of the cloud. Clouds with shorter life times spend \emph{more} time in a state in which $\eta_{\rm GMC}>\epsilon_{\rm GMC}$. 

Our model exemplifies that it is difficult to prove the existence of a time-varying star formation efficiency based on observational quantities such as $\eta_{\rm GMC}$ or $\eta_{\rm ff}$. This is not to say that such a time-dependence does not exist, we merely conclude that current observational evidence for its existence is insufficient.



\acknowledgments

We thank A. Kravtsov, M. Krumholz and N. Murray for helpful comments.

\bibliographystyle{apj}

\begin{thebibliography}{23}
\expandafter\ifx\csname natexlab\endcsname\relax\def\natexlab#1{#1}\fi

\bibitem[{{Ballesteros-Paredes} \& {Hartmann}(2007)}]{2007RMxAA..43..123B}
{Ballesteros-Paredes}, J., \& {Hartmann}, L. 2007, Rev. Mex. Astron. Astrof.,
  43, 123

\bibitem[{{Bonnell} {et~al.}(2010){Bonnell}, {Smith}, {Clark}, \&
  {Bate}}]{2010arXiv1009.1152B}
{Bonnell}, I.~A., {Smith}, R.~J., {Clark}, P.~C., \& {Bate}, M.~R. 2010, ArXiv
  e-prints

\bibitem[{{Elmegreen}(2000)}]{2000ApJ...530..277E}
{Elmegreen}, B.~G. 2000, \apj, 530, 277

\bibitem[{{Elmegreen}(2007)}]{2007ApJ...668.1064E}
---. 2007, \apj, 668, 1064

\bibitem[{{Feldmann} {et~al.}(2010){Feldmann}, {Gnedin}, \&
  {Kravtsov}}]{Feldmann0}
{Feldmann}, R., {Gnedin}, N.~Y., \& {Kravtsov}, A.~V. 2010, in preparation

\bibitem[{{Goldreich} \& {Kwan}(1974)}]{1974ApJ...189..441G}
{Goldreich}, P., \& {Kwan}, J. 1974, \apj, 189, 441

\bibitem[{{Huff} \& {Stahler}(2006)}]{2006ApJ...644..355H}
{Huff}, E.~M., \& {Stahler}, S.~W. 2006, \apj, 644, 355

\bibitem[{{Krumholz} {et~al.}(2006){Krumholz}, {Matzner}, \&
  {McKee}}]{2006ApJ...653..361K}
{Krumholz}, M.~R., {Matzner}, C.~D., \& {McKee}, C.~F. 2006, \apj, 653, 361

\bibitem[{{Krumholz} \& {McKee}(2005)}]{2005ApJ...630..250K}
{Krumholz}, M.~R., \& {McKee}, C.~F. 2005, \apj, 630, 250

\bibitem[{{Krumholz} \& {Tan}(2007)}]{2007ApJ...654..304K}
{Krumholz}, M.~R., \& {Tan}, J.~C. 2007, \apj, 654, 304

\bibitem[{{Lada} {et~al.}(2010){Lada}, {Lombardi}, \&
  {Alves}}]{2010arXiv1009.2985L}
{Lada}, C.~J., {Lombardi}, M., \& {Alves}, J.~F. 2010, ArXiv e-prints

\bibitem[{{Li} \& {Nakamura}(2006)}]{2006ApJ...640L.187L}
{Li}, Z., \& {Nakamura}, F. 2006, \apjl, 640, L187

\bibitem[{{Lopez} {et~al.}(2010){Lopez}, {Krumholz}, {Bolatto}, {Prochaska}, \&
  {Ramirez-Ruiz}}]{2010arXiv1008.2383L}
{Lopez}, L.~A., {Krumholz}, M.~R., {Bolatto}, A.~D., {Prochaska}, J.~X., \&
  {Ramirez-Ruiz}, E. 2010, ArXiv e-prints

\bibitem[{{Matzner}(2002)}]{2002ApJ...566..302M}
{Matzner}, C.~D. 2002, \apj, 566, 302

\bibitem[{{Murray}(2010)}]{2010arXiv1007.3270M}
{Murray}, N. 2010, ArXiv e-prints

\bibitem[{{Murray} {et~al.}(2010){Murray}, {Quataert}, \&
  {Thompson}}]{2010ApJ...709..191M}
{Murray}, N., {Quataert}, E., \& {Thompson}, T.~A. 2010, \apj, 709, 191

\bibitem[{{Myers} {et~al.}(1986){Myers}, {Dame}, {Thaddeus}, {Cohen},
  {Silverberg}, {Dwek}, \& {Hauser}}]{1986ApJ...301..398M}
{Myers}, P.~C., {Dame}, T.~M., {Thaddeus}, P., {Cohen}, R.~S., {Silverberg},
  R.~F., {Dwek}, E., \& {Hauser}, M.~G. 1986, \apj, 301, 398

\bibitem[{{Schruba} {et~al.}(2010){Schruba}, {Leroy}, {Walter}, {Sandstrom}, \&
  {Rosolowsky}}]{2010arXiv1009.1651S}
{Schruba}, A., {Leroy}, A.~K., {Walter}, F., {Sandstrom}, K., \& {Rosolowsky},
  E. 2010, ArXiv e-prints

\bibitem[{{Solomon} {et~al.}(1979){Solomon}, {Sanders}, \&
  {Scoville}}]{1979IAUS...84...35S}
{Solomon}, P.~M., {Sanders}, D.~B., \& {Scoville}, N.~Z. 1979, in IAU
  Symposium, Vol.~84, The Large-Scale Characteristics of the Galaxy, ed.
  {W.~B.~Burton}, 35--52

\bibitem[{{Tan} {et~al.}(2006){Tan}, {Krumholz}, \&
  {McKee}}]{2006ApJ...641L.121T}
{Tan}, J.~C., {Krumholz}, M.~R., \& {McKee}, C.~F. 2006, \apjl, 641, L121

\bibitem[{{V{\'a}zquez-Semadeni} {et~al.}(2005){V{\'a}zquez-Semadeni}, {Kim},
  \& {Ballesteros-Paredes}}]{2005ApJ...630L..49V}
{V{\'a}zquez-Semadeni}, E., {Kim}, J., \& {Ballesteros-Paredes}, J. 2005,
  \apjl, 630, L49

\bibitem[{{Williams} \& {McKee}(1997)}]{1997ApJ...476..166W}
{Williams}, J.~P., \& {McKee}, C.~F. 1997, \apj, 476, 166

\bibitem[{{Zuckerman} \& {Evans}(1974)}]{1974ApJ...192L.149Z}
{Zuckerman}, B., \& {Evans}, II, N.~J. 1974, \apjl, 192, L149

\end{thebibliography}

\clearpage




\end{document}